\documentclass[doublecol]{epl2}                          

\usepackage{amsfonts, amsmath, amssymb}
\usepackage{amscd, latexsym}
\usepackage{mathrsfs}
\usepackage{graphicx} 
\usepackage{epstopdf} 
\usepackage{dcolumn}
\usepackage{bm}
\usepackage{rotating,fancybox}

\title{{Dynamical generation of Floquet Majorana flat bands \\ in s-wave superconductors}}
\shorttitle{Dynamical generation of Floquet Majorana flat bands}

\author{A. Poudel\inst{1} \and G. Ortiz\inst{2} \and L. Viola\inst{1}}
\shortauthor{A. Poudel \etal}

\institute{                    
  \inst{1} {Department of Physics and Astronomy, Dartmouth
               College, 6127 Wilder Laboratory, Hanover, NH 03755, USA}\\
  \inst{2} {Department of Physics, Indiana University,
Bloomington, IN 47405, USA}
}


\abstract{
We present quantum control techniques to engineer flat bands of
symmetry-protected Majorana edge modes in $s$-wave superconductors.
Specifically, we show how periodic control may be employed
for designing time-independent effective Hamiltonians, which support
{\em Floquet Majorana flat bands}, starting from equilibrium conditions that are either 
topologically trivial or only support individual Majorana pairs.  In the first approach, a 
suitable modulation of the chemical potential simultaneously induces Majorana flat bands 
and dynamically \emph{activates} a pre-existing chiral symmetry which is responsible for 
their protection. In the second approach, the application of effective parity kicks  
dynamically {\em generates} a desired chiral symmetry by suppressing 
chirality-breaking terms in the static Hamiltonian. Our results demonstrate how 
the use of time-dependent control enlarges the range of possibilities for realizing gapless 
topological superconductivity, potentially enabling access to topological states of matter 
that have no known equilibrium counterpart.
}

\begin{document}
\maketitle

Characterizing topological phases of matter and the topologically protected 
boundary states that distinguish them is a central challenge across condensed-matter 
physics. Majorana fermions (MFs) in topological superconductors, in particular, carry a 
profound fundamental and practical significance in view of their potential to realize non-Abelian 
exchange statistics and as well as offer topologically protected encodings for 
fault-tolerant quantum computation \cite{Alicea2012}.  While a variety of theoretical proposals 
for model systems supporting MFs have been put forward~\cite{TSproposals,Deng13}, 
and promising experimental signatures in both hybrid superconductor-semiconductor and 
ferromagnetic nanowires~\cite{Exp} are being reported, existing evidence  
of these exotic fermions remains as yet inconclusive~\cite{Liu12}.

Recently, it has been appreciated that additional venues for engineering topological quantum 
matter may become available in systems driven out-of-equilibrium by a suitable 
Hamiltonian~\cite{Kitagawa10,Lindner11} or dissipative~\cite{Dissip} external control action.  In particular, 
in view of rapid experimental progress in optical and microwave coherent-control capabilities of solid-state 
and cold-atom systems, periodic driving provides an especially attractive framework for generating non-equilibrium 
Majorana edge states in both topological superconductors~\cite{Kundu12,Liu13} and superfluids~\cite{ColdAtoms}.
Remarkably, Floquet MFs engineered in this fashion possess distinct transport signatures~\cite{Kundu12} 
and may exist in wider parameter regimes as compared to equilibrium scenarios~\cite{Tong13}.  
Building on these advances, it is both a natural and important question to determine whether 
time-periodic driving may also allow to access non-equilibrium counterparts to surface 
{\em Majorana flat bands} (MFBs), as supported at equilibrium by {\em gapless} topological 
superconductors \cite{You13,sato,Deng14}.  Aside from their intrinsic physical interest, 
gapless topological phases offer unique advantages from a practical standpoint, as the 
presence of a (thermodynamically) large number of MFs potentially results in stronger 
observable signatures essential for accurate detection in two- and three-dimensional 
materials~\cite{Deng14}.

In this work, we provide periodic quantum control protocols for engineering {\em symmetry-protected Floquet 
MFBs} in an $s$-wave superconductor, which may be in a topologically trivial phase at equilibrium.  
We focus on two complementary approaches: on the one hand, we may {\em dynamically activate} 
the chiral symmetry that is responsible for protecting MFBs; on the other hand, we may 
\emph{dynamically generate} a chiral symmetry that is not originally present, and that 
subsequently protects the engineered MFBs.  In the process, we demonstrate how external driving 
can access a richer range of dynamical behavior, including truly non-equilibrium (non-zero-energy)
MFBs and the possibility to evade the anomalous bulk-boundary correspondence (BBC) that known gapless 
topological superconductors exhibit at equilibrium~\cite{Deng14}. 

{\bf Control-theoretic framework.--} We consider a general open-loop control setting described by a time-dependent 
Hamiltonian of the form $H(t) = H_0 + H_c(t)$, where the static Hamiltonian $H_0$ describes the free evolution of the  
system under equilibrium conditions, and $H_c(t)$ represents a suitably chosen control Hamiltonian, driving the 
system out of equilibrium.  While different control protocols will be examined, they share the common requirement of being 
{\em cyclic}, that is, leading to a periodic control propagator, $U_c(t)\equiv {\mathcal T}\exp[-i \int_0^t H_c(t') dt'] 
=U_c(t+T_c)$ for some period $T_c >0$ (in units $\hbar =1$). 

Specifically, we focus on a static Hamiltonian $H_0$ which describes a class of two-band, 
time-reversal invariant topological superconductors on a 2D square lattice in the $\hat{x}$-$\hat{z}$ plane, 
as introduced in our earlier work~\cite{Deng13,Deng14}.  Unlike one-band models which typically require $p$- or $d$-wave 
superconducting pairing~\cite{You13}, our two-band models have the practical advantage of 
using only conventional bulk $s$-wave pairing.  Let ${\bf k} \equiv (k_x,k_z)$ denote the wave-vector 
in the first Brillouin zone.  Then the free momentum-space Hamiltonian in the Nambu basis 
$\psi_{\bf{k}}^\dag \equiv (c_{{\bf{k}},\uparrow}^\dag, c_{{\bf{k}},\downarrow}^\dag,
d_{{\bf{k}},\uparrow}^\dag,d_{{\bf{k}},\downarrow}^\dag,c_{-{\bf{k}},\uparrow},
c_{-{\bf{k}},\downarrow},d_{-{\bf{k}},\uparrow},d_{-{\bf{k}},\downarrow})$ 
may be written as $H_0=\frac{1}{2}\sum_{\bf k} \psi_{\bf{k}}^{\dag} \hat{H}_0({\bf k}) \psi_{\bf{k}} $, 
where 
\begin{align}
\label{Ham}
\hat{H}_0({\bf k})&=s_z (m_{\bf k} \tau_z \hspace*{-0.5mm} - \hspace*{-0.5mm}\mu) + \tau_x 
({\lambda_{k_x}} \sigma_x \hspace*{-0.5mm} + \hspace*{-0.5mm}{\lambda_{k_z}} \sigma_z) + 
\hat{H}_{\text{sw},\zeta}({\bf k})\,. 
\end{align}
Here, the Pauli matrices $s_\nu, \tau_\nu,\sigma_\nu$, $\nu= x,y,z$,  act in the Nambu, orbital, and spin 
space, respectively, and tensor-product notation is understood.  The parameter $\mu$ is the chemical 
potential, whereas $m_{\bf{k}} \equiv {\sf u}_{cd}-2 {\sf w} (\cos{k_x}+\cos{k_z})$, with ${\sf u}_{cd}$ and ${\sf w}$ 
representing the orbital-dependent on-site potential and the intra-band hopping strength. 
Inter-band spin-orbit (SO) interaction is described by ${\lambda_{\bf k}} \equiv (\lambda_{k_x}, \lambda_{k_z})=
-2\,(\lambda_x\,\sin{k_x}, \lambda_z\,\sin{k_z})$, where $\lambda_{x,z}$ are coupling strengths along the 
$\hat{x}$ and $\hat{z}$ directions~\cite{Dimmock66}. The term $\hat{H}_{\text{sw},\zeta}({\bf k})$ 
gives rise to two variants of the above Hamiltonian, $\zeta ={\sf s,t}$, which differ in the nature of the 
(interband, $s$-wave) superconducting pairing: namely, 
$\hat{H}_{\text{sw},{\sf t}}({\bf k}) =-\Delta s_{x} \tau_{y} \sigma_{x}$ for spin-triplet pairing, and 
$\hat{H}_{\text{sw},{\sf s}}({\bf k}) =-\Delta s_{y} \tau_{x} \sigma_{y}$ for spin-singlet pairing, 
where in both cases $\Delta\in {\mathbb R}$ denotes the mean-field superconducting gap~\cite{Pairing}.
For non-vanishing SO coupling (say, $\lambda_x=\lambda_z\equiv \lambda \ne 0$),
the spin-triplet model describes a {\em gapless} $s$-wave superconductor, which supports a 
continuum of Majorana modes in the thermodynamic limit (a MFB) 
for an appropriate choice of boundary conditions.
For generic SO coupling ($\lambda_x, \lambda_z \ne 0$), the spin-singlet model describes instead a 
{\em gapped} $s$-wave superconductor, which only admits a pair of MFs in non-trivial topological phases.   
Remarkably, in this case the system may undergo a topological transition to a gapless superconducting 
phase and MFBs may emerge, provided that the $z$-component of the SO coupling vanishes (i.e., $\lambda_z=0$).

Irrespective of the specific ($\zeta={\sf s,t}$) model Hamiltonian, in {\em all} cases where 
MFBs are found, their existence is protected by a {\em chiral symmetry operator}~\cite{Deng14}, 
namely, a suitable unitary operator ${\mathcal K}_\zeta$ that anti-commutes with ${H}_0$. 
With this in mind, starting from equilibrium situations where MFBs do {\em not} exist, the control problem 
we are interested in is to design a time-independent effective Hamiltonian $H_{\text{eff}}$, which 
approximates the evolution under the time-dependent Hamiltonian $H(t)$, and which supports 
{\em non-equilibrium} MFBs.  As mentioned, two strategies will be employed to ensure that the 
necessary condition $[H_{\text{eff}}, {\mathcal K}_\zeta]_+ =0$ is met: 
if the relevant chiral symmetry is present at equilibrium but dormant, 
$[H_0, {\mathcal K}_\zeta]_+ = 0$, the applied control should steer the dynamics to topologically non-trivial 
regimes for $H_{\text{eff}}$, where MFBs exist and the symmetry protection 
granted by ${\mathcal K}_\zeta$ kicks in; if $[H_0, {\mathcal K}_\zeta]_+ \ne 0$ instead, control 
should symmetrize the dynamics so that chirality-breaking contributions are suppressed and 
protected MFBs may emerge for $H_{\text{eff}}$.

{\bf Floquet formalism.--} 
A general framework for characterizing the effective Hamiltonians that emerge under periodic 
control is provided by Floquet formalism~\cite{Floquet}. The central observation is that the 
evolution under $H(t)$ may be described in terms of basis states of 
the form $\Psi_\alpha (t) \equiv e^{-i \varepsilon_\alpha t} \Phi_\alpha (t)$, where the Floquet 
eigenstates $\Phi_\alpha(t)$ are periodic with period $T_c \equiv 2\pi/\omega$, and the 
Floquet quasi-energies $\varepsilon_\alpha$ are {\em time-independent} but defined only up to 
multiples of $n \omega$, with $n \in {\mathbb Z}$ labeling different ``quasi-energy Brillouin zones''.  
Upon substitution into the Schr\"odinger's equation,
\begin{equation}
\hspace*{-0.5mm}
\Big[{H}(t) - i \frac{\partial}{\partial t}\Big] \hspace*{-0.4mm}
\Phi_\alpha(t) = \varepsilon_\alpha \Phi_\alpha(t) \equiv {H}^{(F)} \Phi_\alpha(t) ,
\label{Floquet}
\end{equation} 
one may formally interpret the original time-dependent problem as a {\em time-independent} eigenvalue 
problem for a {\em Floquet Hamiltonian} operator ${H}^{(F)}$ acting on 
an extended {\em Floquet space} ${\mathcal H}_F \equiv {\mathcal F}\otimes {\mathcal H}$: here, 
${\mathcal H}$ is the physical Hilbert space and the Fourier space ${\mathcal F}$ is the space of all 
$T_c$-periodic functions, ${\mathcal F}= \text{span} \{ e^{i \omega n t} \equiv  | n\rangle, \:  n \in {\mathbb Z}\}$.  
While the Floquet eigenstates form a complete set in ${\mathcal H}_F$, restriction to the first quasi-energy
Brillouin zone ($n=0$) still yields a complete set in ${\cal H}$ and thus a 
physical effective Hamiltonian ${H}_\text{eff} = {H}^{(F)}\vert_{n=0}\equiv {H}_F$.  This effective Hamiltonian  
yields an exact description of the {\em stroboscopic} time-evolution generated by $H(t)$, 
that is, if $U(t)\equiv {\mathcal T}\exp[-i \int_0^t H(t') dt']$ and $t_M= M T_c,M\in {\mathbb N}$, then 
\begin{equation}
U(t_M) = e^{-i H_{\text{eff}} t_M} = [e^{-i H_F T_c}]^M \equiv [U(T_c)]^M, 
\label{strobo}
\end{equation}
where the cycle propagator in the last equality is also referred to as {\em Floquet propagator}.  
Since the latter may be shown to coincide with the one determined within average Hamiltonian 
theory~\cite{Floquet}, analytical approximations to the desired $ {H}_{\text{eff}}$ may be obtained 
by identifying $H_F \equiv \overline{H}$, where $\overline{H} = \sum_{\ell=0}^\infty \overline{H}^{(\ell)}$ 
is the perturbative Magnus series expansion associated to ${H}_c(t)$~\cite{Oteo}. 

As long as the applied periodic control is spatially homogeneous, momentum remains a conserved
quantity of the controlled dynamics, allowing for the Floquet formalism to be lifted to the Nambu 
representation.  That is, by letting 
$H(t)=\frac{1}{2}\sum_{\bf k} \psi_{\bf{k}}^{\dag} \hat{H}({\bf k}, t) \psi_{\bf{k}}$, we may write 
$\hat{H}({\bf k}, t) = \hat{H}_0({\bf k}) + \hat{H}_c({\bf k}, t)$ and,   
correspondingly, $H_F=\frac{1}{2}\sum_{\bf k} \psi_{\bf{k}}^{\dag} \hat{H}_F({\bf k}) \psi_{\bf{k}}$. 
Topological features are then encoded in the {\em Floquet quasi-energy spectrum} 
$\{ \varepsilon_0 ({\bf k}) \}$~\cite{Kitagawa10} which, in practical situations, 
is determined numerically.
Depending on the details of the control protocol, this may be more easily carried out in the 
frequency or time domain.  Let the controlled single-particle Hamiltonian be expressed as 
$\hat{H}({\bf k}, t) \equiv \sum_{n=-\infty}^{\infty} \hat{H}^{(n)}({\bf k})\, e^{i n \omega t}$ 
in the extended space ${\mathcal H}_F$.  Then $\hat{H}^{(F)}({\bf k})$ may be constructed as
\begin{align}
\hat{H}^{(F)}({\bf k}) = \sum_{n=-\infty}^{\infty} [Q_n \otimes\hat{H}^{(n)}({\bf k})] + \omega \, 
(Q_z \otimes {\mathbb I})\,,
\label{floq}
\end{align}
where the operators $Q_{z}, Q_n$ act on ${\cal F}$ as $Q_z|m\rangle = m |m\rangle$ and 
$Q_n |m\rangle = |n+m\rangle$, respectively, and ${\mathbb I}$ denotes the identity operator 
on ${\mathcal H}$. The desired quasi-energies are then obtained by block-diagonalizing a 
suitably truncated version of the above matrix and by restricting to $n=0$ afterwards.
Alternatively, since the Floquet eigenstates are also eigenstates of the 
Floquet propagator $\hat{U}_F ({\bf k}) = \hat{U} ({\bf k}, T_c)  = 
\mathcal{T}\mbox{exp}[-i \int_0^{T_c} \hat{H}({\bf k}, t') dt']$, that is, 
\begin{equation}
\hat{U}_F ({\bf k}) \Phi_\alpha({\bf k}, T_c) =
e^{-i \varepsilon_\alpha ({\bf k})\,T_c } \Phi_\alpha({\bf k}, T_c) , 
\label{F2}
\end{equation}
one may directly extract the desired quasi-energy spectrum  
by computing and diagonalizing $\hat{U}({\bf k}, T_c)$. 

{\bf Floquet Majorana flat bands via dynamical chiral-symmetry activation.--}
We first consider a static Hamiltonian describing an $s$-wave gapless spin-triplet superconductor,
according to Eq. (\ref{Ham}) with $\zeta={\sf t}$. In this case, $[\hat{H}_0({\bf k}), U_{K_{\sf t}} ]_+=0$~\cite{Deng14}, 
where the unitary operator $U_{K_{\sf t}}$ describing the action of the chiral symmetry operator ${\cal K}_{\sf t}$
in the Nambu space has the explicit form $U_{K_{\sf t}} = s_x \otimes \tau_z \otimes I$, with $I$ 
denoting the $2\times2$ identity matrix~\cite{Chiral}.
Control is introduced via periodic modulation of the chemical potential in $\hat{H}_0({\bf k})$, 
namely, $\hat{H}_c({\bf k}, t)= \mu_d\,\cos(\omega t)\,s_z \otimes I \otimes I$, 
where $\mu_d$ and $\omega = 2\pi/T_c$ are the driving amplitude and frequency, respectively.
Note that the time dependence of the chosen driving protocol is such that the 
Floquet Hamiltonian satisfies $[\hat{H}_F({\bf k}), U_{K_{\sf t}} ]_+=0$.  To establish this, 
in view of the stroboscopic equality $\hat{H}_F ({\bf k})= \overline{H}({\bf k})$, it suffices to show 
that $[\overline{H}({\bf k}), U_{K_{\sf t}} ]_+=0$. By using the fact that chirality is preserved by the 
{\em instantaneous} time-dependent Hamiltonian $\hat{H}({\bf k}, t)$, 
direct calculation shows that any chirality-breaking term in the Magnus expansion vanishes. 
The presence of chiral symmetry does not guarantee, however, the existence of MFBs.  
We now demonstrate the possibility to access control regimes for which the Floquet 
quasi-energy spectrum does support non-equilibrium MFBs, starting from a trivial 
topological phase at equilibrium. 

\begin{figure}[t] 
\begin{centering}
\hspace*{5mm}\includegraphics[width=8.0cm]{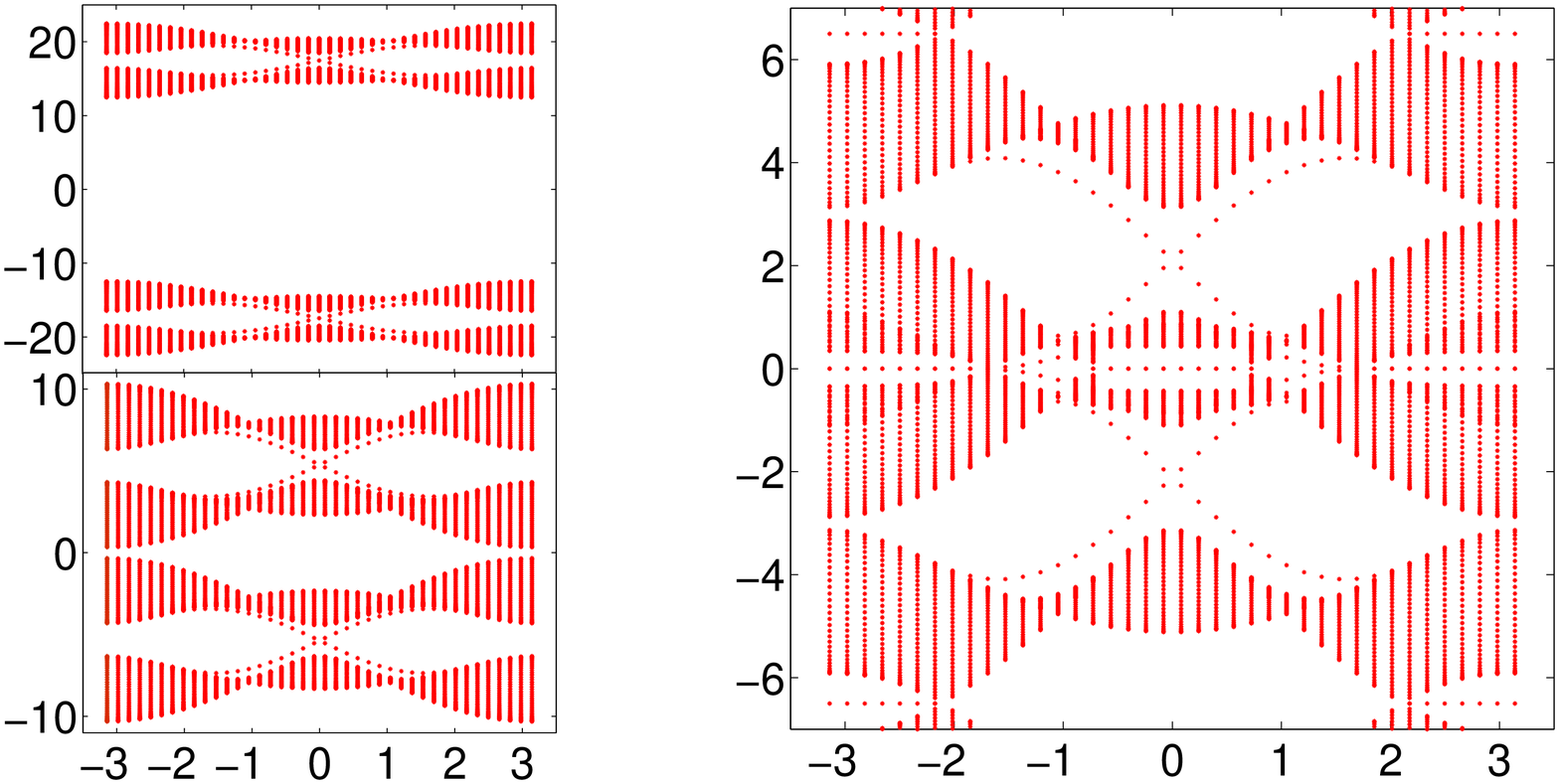}  
\put(-238, 65){\rotatebox{90}{$\epsilon(k_z)$}}
\put(-132, 55){\rotatebox{90}{$\varepsilon_0(k_z)$}} 
\put(-185, 2){$k_z$}
\put(-60, 2){$k_z$}
\end{centering}
\vspace*{-3mm} 
\caption{(Color online) Left panels: single-particle spectra $\epsilon({k_z})$ of the spin-triplet static 
Hamiltonian $\hat{H}_0(k_z)$ for $\lambda = {\sf w} = {\sf u}_{cd} = 1$, $\Delta = 4$, and $\mu = 3$ (bottom), 
$\mu = 17$ (top). Right panel:  Floquet quasi-energy spectrum $\varepsilon_0(k_z)$ for the same geometry 
and boundary conditions, with driving amplitude and frequency $\mu_d = 7$, $\omega = 13$. 
Here and, unless otherwise stated in the rest of the paper, the lattice size is 
$(N_x, N_z)=(80, 80)$, and PBC (OBC) are assumed in the $\hat{z}$ ($\hat{x}$) direction, respectively. 
The quasi-energy spectra reveal MFs at zero as well as $\pm\omega/2$ energies, 
while the static counterpart does not support any MFs.}
\label{f1}
\end{figure}

The plots in the left panel of Fig.~\ref{f1} depict energy spectra of the static Hamiltonian $\hat{H}_0({\bf k})$ 
in a strip geometry with periodic boundary conditions (PBC) in the $\hat{z}$ direction and open boundary 
conditions (OBC) in the $\hat{x}$ direction. For the chosen set of parameters (in fact, for {\em any} value 
of the chemical potential $\mu \in [3, 17]$) the system is in a topologically trivial phase at equilibrium, and no 
zero-energy mode exists. The quasi-energy spectrum of the driven system, computed up to the $5$-{th} 
harmonic of $\hat{H}^{(F)}({\bf k})$~\cite{Floquet5}, is shown in the right panel of Fig.~\ref{f1} 
upon restriction to the first quasi-energy Brillouin zone, $-\omega/2\leq \varepsilon_0 (k_z) < \omega/2$.
In the limit $\mu_d \rightarrow 0$, the Floquet spectrum consists of an array of copies of the original energy 
spectrum of $\hat{H}_0(k_z)$, shifted up and down by an integer multiple of $\omega$. For $\mu_d \ne 0$, the 
driving induces hopping between levels with different $n$ values. Due to the monochromatic nature of the driving, 
$n$ can differ at most by one. Physically, the increase (decrease) of $n$ may be 
interpreted as absorption (emission) of a photon from the driving field. 
Upon increasing $\mu_d$, two overlapping bands whose frequency index $n$ differs by one start to fold back 
forming a gap (akin to avoided-crossing in a driven two-level system), first around $\pm \omega/2$ energies 
and, for larger $\mu_d$, around zero energy as well. It is in between this gap that the Floquet MFBs emerge.
Seemingly counterintuitively, since $\mu$ never exceeds the range $\mu \in [3,17]$, {\em the instantaneous 
Hamiltonian remains in a topologically trivial phase} throughout the duration of the driving.  Notwithstanding, the 
Floquet spectrum reveals flat bands of zero-energy modes, as well additional flat bands at energies 
$\pm \omega/2$. The periodic nature of the quasi-energies, together with the particle-hole symmetry of the 
Bogoliubov-de Gennes Hamiltonian, which is transferred to $\hat{H}_F({\bf k})$, cause states with 
$\pm \omega/2$ energies to become their own particle-hole conjugates, thus satisfying the defining Majorana property. 
These non-zero-energy MFBs lack a static counterpart, and thus truly consist of non-equilibrium Majorana 
edge modes~\cite{Remark}. 
 
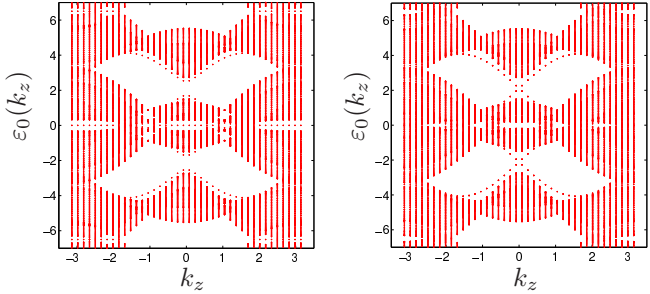
\begin{figure} 
\includegraphics[width=8.5cm]{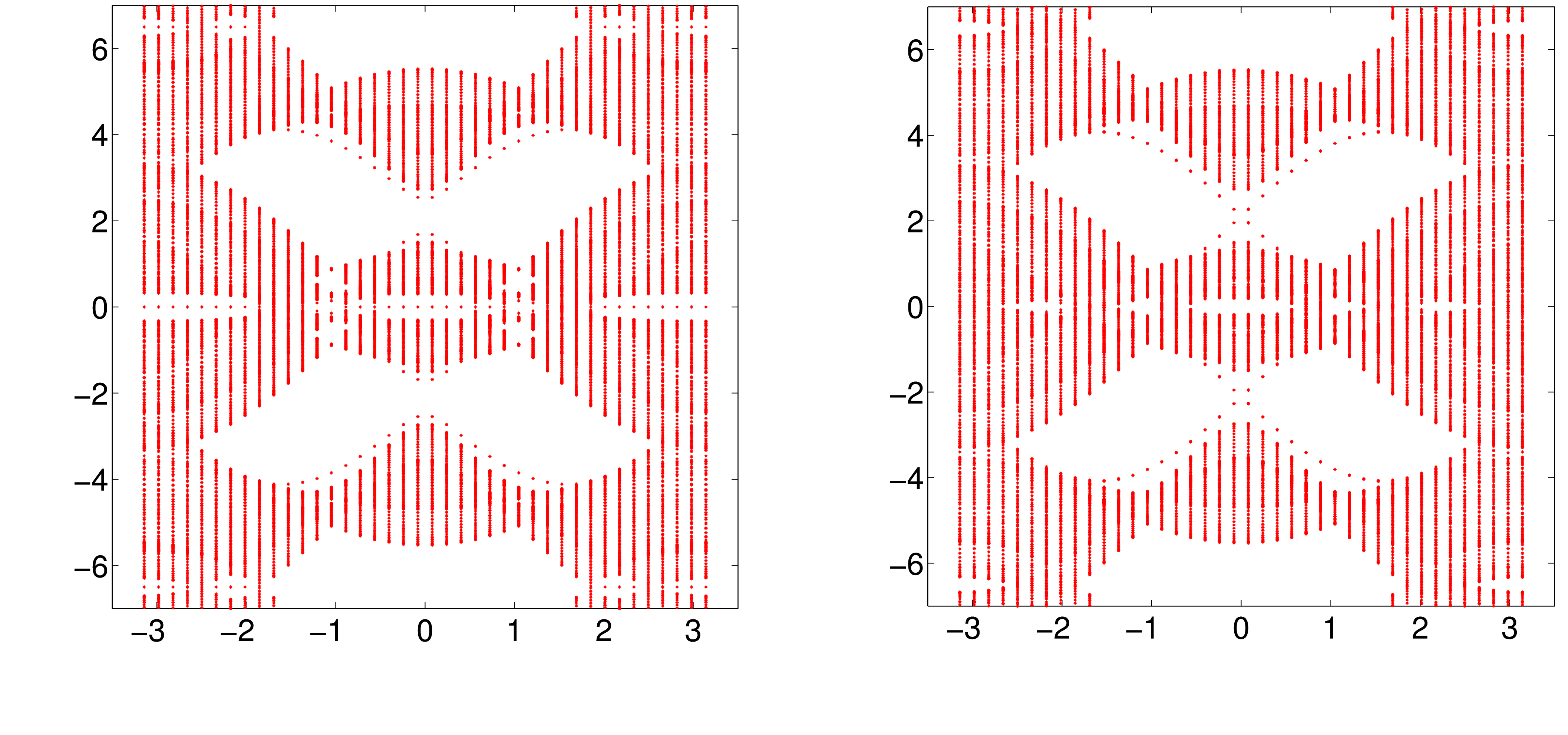} \vspace*{-3mm} 
\put(-248, 60){\rotatebox{90}{$\varepsilon_0(k_z)$}}
\put(-120, 60){\rotatebox{90}{$\varepsilon_0(k_z)$}} 
\put(-182, 5){$k_z$}
\put(-56, 5){$k_z$}
\caption{(Color online) Floquet quasi-energy spectra 
in the presence of a static magnetic field along the $\hat{x}$ direction, with $h_x = 0.5$ (left panel), and along the $\hat{y}$ direction, with $h_y=0.5$ (right panel). All other parameters are the same as in the right panel of Fig.~\ref{f1}. The anti-commutation (commutation) property of the field in the $\hat{x}$ ($\hat{y}$) direction with the \emph{activated} chiral symmetry operator $\mathcal{K}_t$ leads to protection (destruction) of the Floquet MFBs.}
\label{f2}
\end{figure}

We demonstrate that the \emph{activated} chiral symmetry indeed protects the engineered Floquet MFBs by considering 
the effect of a static magnetic field, described by an additional Hamiltonian 
$H_\nu=h_\nu \sum_j \psi_j^\dag \sigma_\nu \psi_j$, with $\nu={x},{z}$ $({y})$ 
corresponding to in-plane (out-of-plane) directions. In the Nambu space, the perturbing Hamiltonian reads 
$\hat{H}_{\nu}({\bf k}) = h_{\nu} \, s_z(I) \otimes I \otimes \sigma_\nu$ for in-plane (out-of-plane) directions. 
For an in-plane field, the Floquet MFBs remain essentially flat, Fig.~\ref{f2}(a). However, if $h_y\ne 0$, the 
Floquet MFBs become unstable, Fig.~\ref{f2}(b), and eventually cease to exist. The effect of the magnetic field 
along different directions on the Floquet MFBs is similar to its effect on equilibrium MFBs, and can be understood 
in terms of chiral-symmetry protection. Specifically, the $\hat{x}$ and $\hat{z}$-components of $\hat{H}_\nu({\bf k})$ 
anti-commute with $U_{K_{\sf t}}$, whereas $\hat{H}_y({\bf k})$ {\em commutes} with $U_{K_{\sf t}}$.  Accordingly, 
chirality-protection of the Floquet MFBs is lost in this case, just as it is in the static case.

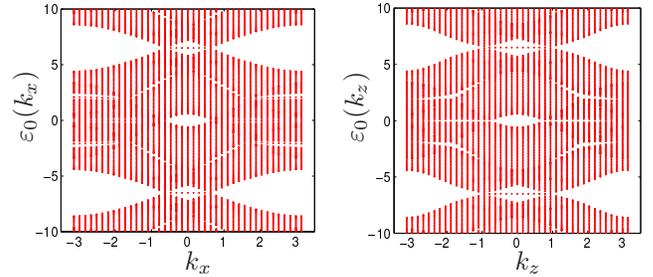
\begin{figure}[t] 
\includegraphics[width=8.5cm]{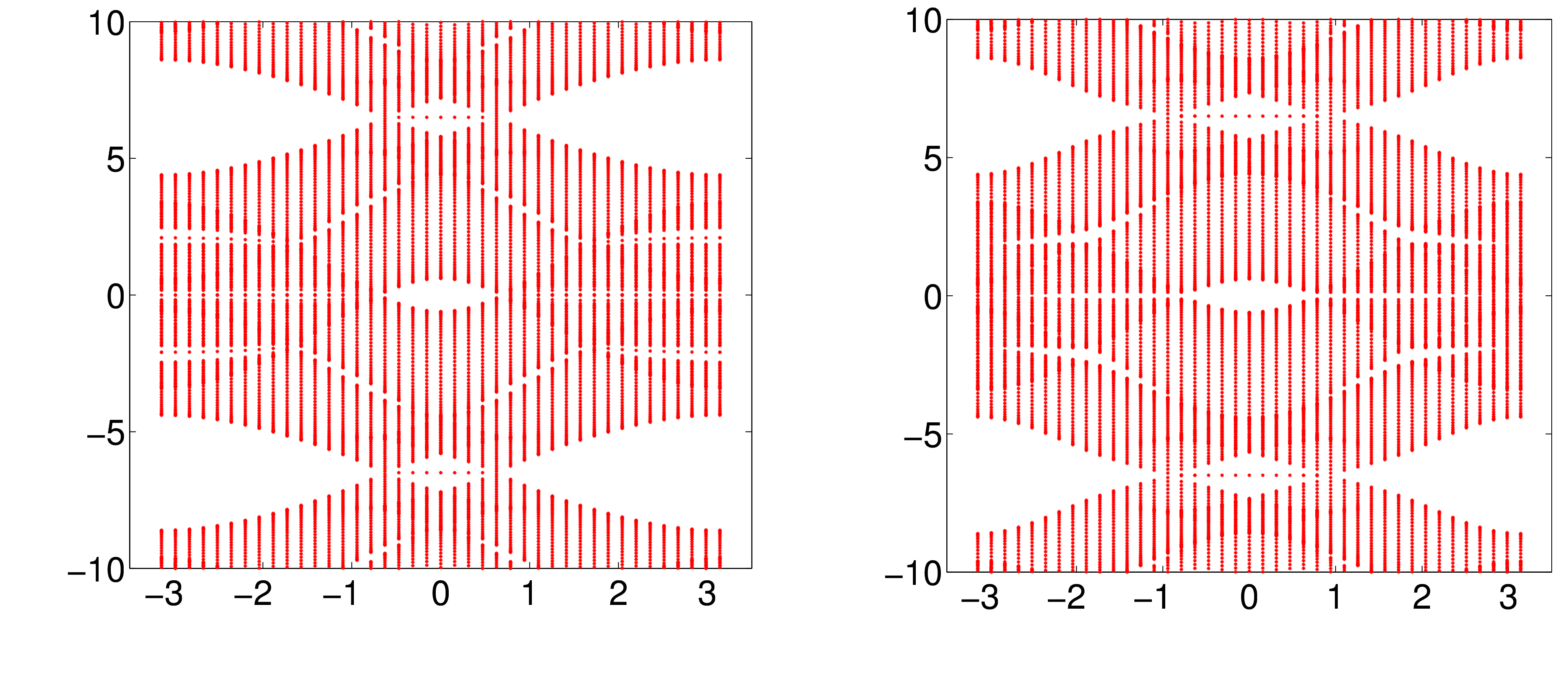} 
\put(-240, 50){\rotatebox{90}{$\varepsilon_0(k_x)$}}
\put(-115, 50){\rotatebox{90}{$\varepsilon_0(k_z)$}} 
\put(-175, 2){$k_x$}
\put(-50, 2){$k_z$}
\vspace*{-3mm} 
\caption{(Color online) Floquet quasi-energy spectrum in a strip geometry under periodic modulation of 
the $\hat{x}$-component of the magnetic field. Left panel: PBC in the $\hat{z}$ direction 
and OBC in the $\hat{x}$ direction.  Right panel: PBC and OBC are interchanged. 
The parameters are ${\sf w} = 1, {\sf u}_{cd} = 4, \Delta = 2, h_{x_0} = 8, \mu = 0$. The magnetic driving 
amplitude and frequency are $h_x=7$ and $\omega=13$. The controlled system admits Floquet MFBs at 
energies $\pm \omega/2$, {\em independent} of the choice of OBC, unlike the equilibrium case or when 
$\mu$ is modulated.}
\label{f3}
\end{figure}

As our previous work on MFBs revealed, gapless topological superconductivity at equilibrium 
is distinguished by an anomalous BBC, as compared to the gapped case~\cite{Deng14}. 
Namely, the emergence of MFBs depends in general on the direction along which OBC vs PBC are assigned, 
reflecting an asymmetric momentum dependence of the quasi-particle excitation gap.
We find that such an anomalous BBC may persist for Floquet MFBs as well. Specifically, in the driven model 
we have considered thus far, we have verified (data not shown) that {\em no} Floquet MFBs exist 
if we assign OBC in the $\hat{z}$ direction -- precisely as in the equilibrium scenario.
However, this anomalous BBC may be eliminated by suitably modifying the control protocol. In our case, 
we replace the periodic modulation of the chemical potential with a periodic in-plane magnetic field, 
that is, we consider a control protocol described by 
$\hat{H}_c ({\bf k}, t) = [h_{x_0} + h_x  \cos(\omega t)]\, s_x \otimes I \otimes \sigma_z$. In view of the 
above observation, chirality is preserved at all times by the resulting time-dependent Hamiltonian $H(t)$.
The Floquet quasi-energy spectrum under periodic magnetic driving is shown in Fig.~\ref{f3} for the two 
possible choices of boundary conditions, where again we choose parameters such that initially the 
system is topologically trivial. Surprisingly, the applied control eliminates the anomalous BBC and allows 
for Floquet MFBs to exist {\em along both boundaries}, albeit at non-zero $\pm \omega/2$ energies only (at 
least for relatively weak field amplitudes as we consider). Even more remarkably, this driven model provides, 
to the best of our knowledge, the only example of an {\em $s$-wave} gapless topological superconductor 
supporting MFBs in {\em both} directions, with other known models requiring $d_{xy}$-wave pairing~\cite{sato}.  
This shows explicitly how driven quantum matter may access a broader range of possibilities, 
including non-equilibrium quantum phases without known equilibrium counterpart.

{\bf Floquet Majorana flat bands via chiral-symmetry generation.--}
We now turn to a static Hamiltonian describing an $s$-wave {\em gapped spin-singlet} superconductor, 
according to Eq. (\ref{Ham}) with $\zeta={\sf s}$. As mentioned, MFBs may exist in equilibrium only if the 
$z$-component of the SO coupling vanishes, $\lambda_z=0$. If so, the MFBs are protected by a chiral 
symmetry operator ${\cal K}_{\sf s}$, whose unitary action in Nambu space is described by 
$U_{K_{\sf s}} = s_z \otimes \tau_y \otimes \sigma_x$~\cite{Chiral}. However, any non-zero $\lambda_z$ tilts 
the MFBs into a pair of MFs along each boundary, reflecting the fact that the $z$-component of the SO 
coupling {\em commutes} with ${\cal K}_{\sf s}$.  In situations where  $\lambda_z$ is not directly tunable, 
we show how one may in principle suppress its effect dynamically, by using periodic control. 

We consider a special instance of a dynamical decoupling protocol also known as ``parity-kick'' 
method, whose net effect may be thought of as dynamically enforcing a desired symmetry 
on the effective dynamics~\cite{Viola98}. The idea is to devise a control Hamiltonian $H_c(t)$ that 
implements a sequence of instantaneous, uniformly spaced kicks, with each control cycle involving two 
kicks and with each kick mimicking the (unitary) parity operator for the system, say ${\mathcal P}$. 
Let $H_{\pm}$ denote the static Hamiltonian ${H}_0$ that corresponds to Eq. (\ref{Ham}) 
with $\zeta= {\sf s}$ and $\pm \lambda_z$.  Then we require that ${\cal P}$ flips the sign of $\lambda_z$,
by commuting with all the remaining terms of ${H}_0$:
\begin{equation}
{\cal P}^{-1} \,H_{+} \,{\cal P} = H_{-}, \quad {\cal P}^2= I, 
\label{kick}
\end{equation}
with ${\cal P} \equiv e^{-i \varphi H_{\text{kick}}}$, $H_c (t) = H_{\text{kick}}\sum_{p=1}^\infty \delta(t - p T_c/2)$,  
for a kick Hamiltonian $H_{\text{kick}}$ and a strength parameter $\varphi$ to be determined.   
Using Eq. (\ref{kick}), the single-cycle propagator reads 
$U(T_c) = {\cal P}\, U_{+}(T_c/2)\, {\cal P}\, U_{+}(T_c/2) = e^{-iH_{-} T_c/2}  e^{-iH_{+}T_c/2} 
\equiv e^{- i H_{\text{eff}}T_c}$. Up to the second order in the Magnus series, $H_{\text{eff}}$
may then be approximated as
\begin{align}
H_{\text{eff}} &\approx \overline{H}^{(2)} = \frac{1}{2} (H_{-} + H_{+})  - \frac{i T_c}{8} \big[H_{-}, H_{+}\big] 
\label{secondHeff} \\  
&-\frac{T_c^2}{96} \Big( [H_{-}, \big[H_{-}, H_{+}]] +[[H_{-}, H_{+}\big], H_{+}] \Big) 
+ {\mathcal O} (T_c^3). \nonumber 
\end{align}
While formal conditions for convergence of the Magnus expansion over time $t_M$ may be stated in 
terms of appropriate norms of the unwanted interaction~\cite{Oteo} 
(loosely, $\vert\vert H_{\text{so},z}\vert\vert t_M <\pi$ in our case, translating into $2\lambda_z t_M 
\lesssim \pi$), these conditions need not be necessary and may in fact be too conservative in situations 
of interest~\cite{Slava,DDremark}.
We also note that the choice $U_P = s_y \otimes \tau_y \otimes \sigma_z$ achieves 
the desired transformation Eq. (\ref{kick}) in the Nambu representation.   An explicit form for the kick 
Hamiltonian and strength parameter $H_{\text{kick}}, \varphi$ may be also obtained, upon 
exploiting the action of ${\cal P}$ on the fermionic operators~\cite{ParityKick}. 

\begin{figure}[t] 
\includegraphics[width=8.5cm]{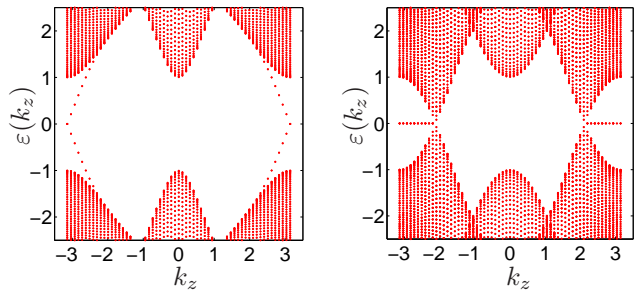} 
\vspace*{-3mm} 
\put(-244, 56){\rotatebox{90}{$\varepsilon(k_z)$}}
\put(-118, 56){\rotatebox{90}{$\varepsilon(k_z)$}} 
\put(-182, 3){$k_z$}
\put(-56, 3){$k_z$}
\caption{(Color online) (Quasi)-energy spectra $\varepsilon(k_z)$ of the second-order effective 
Hamiltonian of Eq.~(\ref{secondHeff}) in a strip geometry with PBC (OBC) in the $\hat{z}$ ($\hat{x}$) direction, and  
${\sf w} = {\sf u}_{cd} = \lambda =1, \Delta = 4, \mu = 0$. Left panel: time $t=0$. Right panel: $t=T_c=0.1$. 
The pair of equilibrium MFs is ``deformed'' into non-equilibrium MFBs in the presence of periodic kicks.}
\label{f4}
\end{figure}

The quasi-energy spectrum of the second-order effective Hamiltonian $\hat{H}_{\text{eff}}({\bf k})$
resulting from Eq. (\ref{secondHeff}) is shown in Fig.~\ref{f4}, in the usual strip geometry. 
As seen in the left panel, in the absence of control only a pair of MFs is present on each boundary for 
the chosen equilibrium parameters. However, upon driving the system, MFBs clearly emerge, 
as shown in the right panel of Fig.~\ref{f4}.  By construction, the unwanted $\hat{z}$-component of the 
SO coupling is exactly canceled to ${\mathcal O}(T_c^0)$ in Eq. (\ref{secondHeff}). 
However, terms proportional to $\lambda_z$ may potentially be re-introduced to higher order. 
Remarkably, using the property ${\cal K}_{\sf s}\,H_{\pm}\,{\cal K}_{\sf s}^{-1} = -H_{\mp}$, it follows 
that ${\cal K}_{\sf s}\,e^{- iH_{\text{eff}}T_c}\,{\cal K}_{\sf s}^{-1} =
[{\cal K}_{\sf s}\,e^{-i H_{-}T_c/2}\,{\cal K}_{\sf s}^{-1}] [{\cal K}_{\sf s}\,e^{-iH_{+}T_c/2}\,{\cal K}_{\sf s}^{-1}] 
= e^{+ iH_{\text{eff}}T_c}$, implying that $H_{\text{eff}}$ anti-commutes with the chiral 
symmetry operator {\em up to arbitrary order}.  Thus, MFBs are stable against higher-order Magnus 
corrections, whose main effect (assuming convergence, data not shown) 
is to perturb the overall band structure. Altogether, this shows how the applied control {\em dynamically 
generates} at once Floquet MFBs and the chiral symmetry that protects them. 

\begin{figure} 
\vspace*{-3mm}
\includegraphics[width=8.6cm]{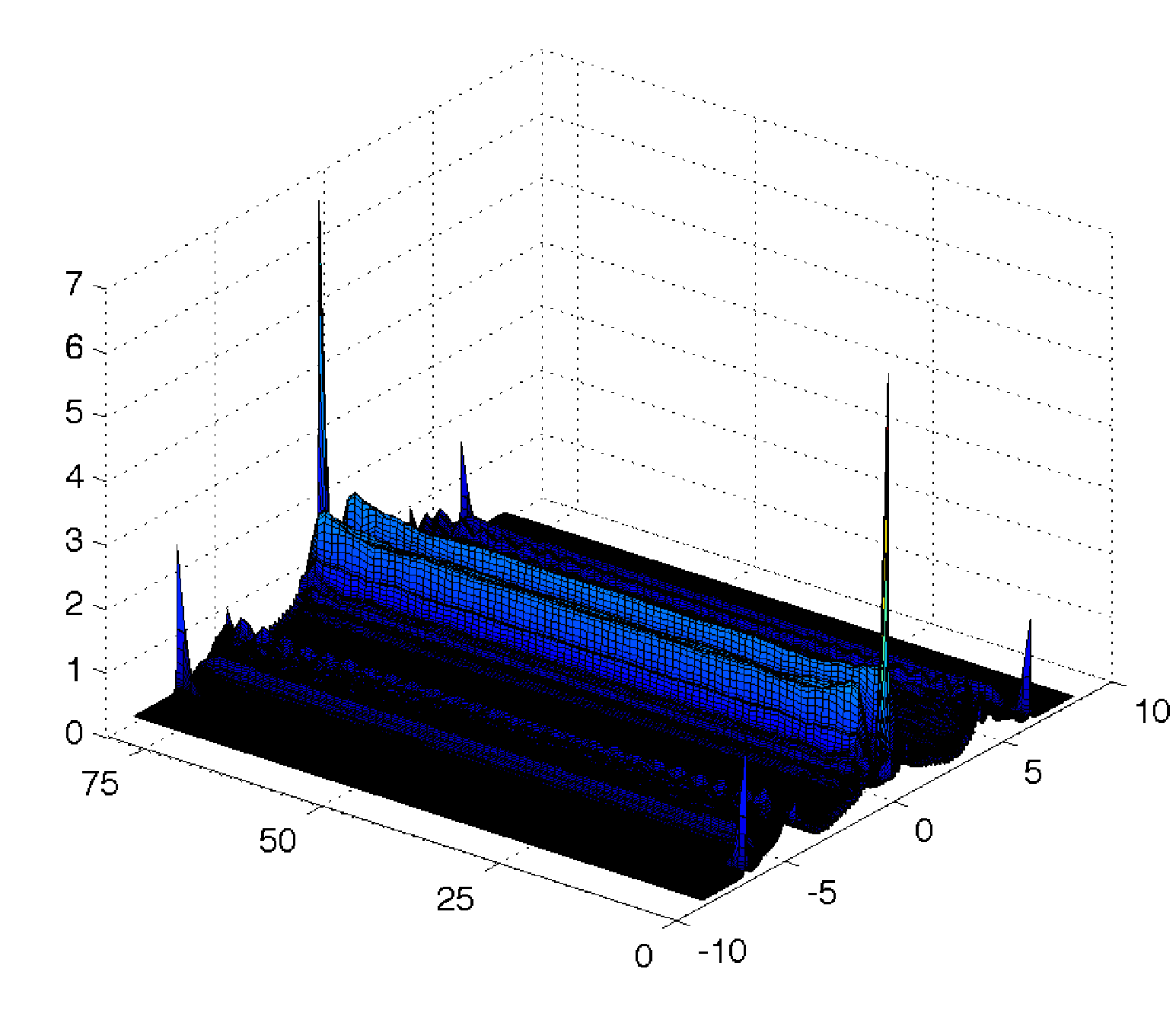} 
\put(-242, 95){\rotatebox{90}{LDOS}}
\put(-45, 30){$\varepsilon$}
\put(-178, 24){$x_i$}
\vspace*{-5mm} 
\caption{(Color online) Quasi-electronic LDOS as a function of quasi-energy $\varepsilon$ 
and lattice site $x_i$ in a strip geometry with PBC (OBC) in the $\hat{z}$ ($\hat{x}$) direction, 
for the same driven $s$-wave spin-triplet superconductor considered in Fig.~\ref{f1} (right panel). The 
LDOS is peaked around quasi-energies $0,\, \pm \omega/2$, consistent with the quasi-energy spectrum 
in Fig.~\ref{f1}. }
\label{f5}
\end{figure}

{\bf Detection of Floquet Majorana flat bands.--}  As stressed in our previous work~\cite{Deng14}, MFBs 
offer distinctive advantages as compared to MF pairs for detection via scanning tunneling microscope 
conductance experiments.   Importantly, these advantages are maintained for non-equilibrium MFBs. 
For a periodically driven material, the (stroboscopic) tunneling current is proportional to the quasi-electronic 
local density of states (LDOS), evaluated from the quasi-energy spectrum~\cite{Kohler05}. 
Thus, the quasi-electronic LDOS may be used as an observable signature for non-equilibrium 
MFBs.  In Fig.~\ref{f5}, we plot the quasi-electronic LDOS of the Floquet Hamiltonian describing 
the driven gapless spin-triplet superconductor that we initially examined. The quasi-electronic LDOS is strongly 
peaked around zero energy, consistent with the quasi-energy spectrum in Fig.~\ref{f1}. Furthermore, additional 
peaks are present around $\pm \omega/2$ on both ends, which conforms with the 
non-zero-energy Floquet MFBs observed in Fig.~\ref{f1}. This plot also serves to verify that Floquet MFBs 
are localized at the boundaries, as expected.

{\bf Conclusions.--} We have shown how to dynamically engineer symmetry-protected Floquet Majorana 
flat bands by applying periodic control to a class of two-band $s$-wave superconductors, starting from 
equilibrium conditions where none or at most a single pair of Majorana edge modes exist.   While, as 
remarked in our previous work~\cite{Deng14}, the special form of the spin-orbit interaction~\cite{Dimmock66} 
makes lead chalcogenide materials particularly promising for experimental realizations of our proposal, 
we stress that our control strategies have potentially far broader applicability. 
For example, in conjunction with synthetic spin-orbit couplings~\cite{Ian}, they may 
expand the toolbox for engineering topological states of matter in ultracold atomic systems.  
Likewise, an interesting extension under current investigation is the possibility  to engineer Floquet flat bands 
of spin-polarized edge modes in topological insulators, which may yield enhanced current in 
spin-torque experiments~\cite{Mellnik14}, with implications for spintronics devices. 

\acknowledgments
\noindent 
L.V. gratefully acknowledges partial support from the NSF through Grant No.
PHY-1104403 and from the Constance and Walter Burke
Special Projects Fund in {\em Quantum Information Science}.


\begin{thebibliography}{100}
\bibitem{Pairing}
The real-space Hamiltonian for triplet pairing is 
$H_{\text{sw},{\sf t}} = i\Delta \sum_j [ (c^\dag_{j,\uparrow} d^\dag_{j,\downarrow} + c^\dag_{j,\downarrow} 
d^\dag_{j,\uparrow}) + \text{H.c.}]$, whereas 
$H_{\text{sw},{\sf s}} = \Delta \sum_j [ (c^\dag_{j,\uparrow} d^\dag_{j,\downarrow} - c^\dag_{j,\downarrow} 
d^\dag_{j,\uparrow}) + \text{H.c.}]$ for singlet pairing.
\bibitem{Remark} Extensive numerical tests indicate that the existence of non-equilibrium MFBs is 
as stable against small parameter variations as in the static equilibrium problem.
\end{thebibliography}
\end{document}